\documentclass[twocolumns,journal]{IEEEtran}
\ifCLASSINFOpdf
\else
\fi
\usepackage{amssymb}
\usepackage{graphicx}
\usepackage{graphics}
\usepackage{color}
\usepackage{mathrsfs}
\usepackage{algorithm}
\usepackage{algorithmic}
\usepackage{amsfonts}
\usepackage{pifont}

\begin{document}
%
\title{Simulation Results of Center-Manifold-Based Identification of Polynomial Nonlinear Systems with Uncontrollable Linearization}
%
%
%

\author{Chao~Huang, Hao~Zhang, Zhuping~Wang 
\thanks{*This work was supported by the
National Natural Science Foundation of China under Grant 62373282, Grant 62350003 and Grant 62150026.}
\thanks{C. Huang, H. Zhang and Z. Wang are with the School of Electronic and Information Engineering, Tongji University, Shanghai 200092, People's Republic of China (Corresponding Author: Chao Huang, E-mail: csehuangchao@tongji.edu.cn).}

}

\maketitle

\begin{abstract}
Recently, a system identification method based on center manifold is proposed to identify polynomial nonlinear systems with uncontrollable linearization. This note presents a numerical example to show the effectiveness of this method.  
\end{abstract}


\newtheorem{Def}{Definition}
\newtheorem{Asu}{Assumption}
\newtheorem{Thm}{Theorem}
\newtheorem{Cor}{Corollary}
\newtheorem{Alg}{Algorithm}
\newtheorem{Rem}{Remark}
\newtheorem{Lem}{Lemma}
\newtheorem{Pro}{Problem}
\newtheorem{pot}{Proof of Theorem}

%
\IEEEpeerreviewmaketitle

\section{Introduction}

Recently, a system identification (SID) method based on center manifold (CM) is proposed to identify polynomial nonlinear systems with \emph{uncontrollable linearization} \cite{Huang2024Analytical}. This note presents some simulation results of the method to show its effectiveness. To make the note self-contained, in Section II, we state the problem for which the CM-based SID method is developed, and in Section III we describe the fairly standard frequency-domain subspace (FDS) algorithm which is the foundation of the CM-based method. Then, Section IV provides the simulation results, which is the main content of the note. Please be aware that in Section IV, each step of the SID algorithm is described, but the principles or reasons behind each step will not be explained. Readers can refer to \cite{Huang2024Analytical} which contains the complete theory for details. In addition, this note assumes that the readers are familiar with the successive Kronecker product and its variant, denoted by $v^{\left(i\right)}$ and $v^{\left[i\right]}$ respectively, the successive Kronecker sum and its variant, denoted by $A^{\left\{i\right\}}$ and $A^{\left\langle i \right\rangle }$ respectively, and their relations such as $v^{\left[i\right]}=M_i^nv^{\left(i\right)}$, $v^{\left(i\right)}=N_i^nv^{\left[i\right]}$ and $A^{\left\langle i \right\rangle }=M_i^nA^{\left\{i\right\}}N_i^n$, where $M_i^n$ and $N_i^n$ are constant matrices with appropriate dimensions, otherwise the readers can refer to e.g., \cite{Huang2024Identification} for detailed background knowledge.

\section{Problem Statement}

Consider the following multi-input-multi-output polynomial nonlinear system (PNS), possibly with an uncontrollable linearization:
\begin{equation}\label{eq_preliminary_nldequ_x0}
\begin{array}{l}
\dot x\left( t \right) = f\left( {x\left( t \right),u\left( t \right)} \right),\\
y\left( t \right) = h\left( {x\left( t \right),u\left( t \right)} \right),\\
\end{array}
\end{equation}
where $x\in\mathbb R^n$ is the system state, $u\in\mathbb R^m$ is the input, and $y\in\mathbb R^p$ is the output, $f:\mathbb R^n\times\mathbb R^m\to\mathbb R^n$ and $h:\mathbb R^n\times\mathbb R^m\to\mathbb R^p$ are functions of polynomial nonlinearity, i.e.,
\begin{equation}\label{eq_f}
f\left( {x,u} \right) = \sum\nolimits_{l = 1}^L {\sum\nolimits_{\scriptstyle i + r = l,\hfill\atop
\scriptstyle i,r \in\mathbb N\hfill} {{{\mathbf{F}}_{i,r}}} \left( {{x^{\left[ i \right]}} \otimes {u^{\left[ r \right]}}} \right)} ,
\end{equation}
\begin{equation}\label{eq_h}
h\left( {x,u} \right) = \sum\nolimits_{l = 1}^L {\sum\nolimits_{\scriptstyle i + r = l,\hfill\atop
\scriptstyle i,r \in\mathbb N\hfill} {{{\mathbf{H}}_{i,r}}} \left( {{x^{\left[ i \right]}} \otimes {u^{\left[ r \right]}}} \right)} ,
\end{equation}
where $L\in\mathbb N^+$ is a known upper bound of the order of nonlinearity, ${\mathbf{F}}_{i,r}\in\mathbb R^{n\times s_{i,r}}$ and ${\mathbf{H}}_{i,r}\in\mathbb R^{p\times s_{i,r}}$ are fixed yet unknown matrices to be identified, where according to \cite{Huang2024Identification}, ${s_{i,r}} = \mathcal C_{n + i - 1}^i\mathcal C_{m + r - 1}^r$, $\mathcal C$ represents the combinatorial operation. Following the notational convention for linear systems, we denote
\[A = {\mathbf F}_{1,0},B = {\mathbf F}_{0,1},C =  {\mathbf H}_{1,0},D =  {\mathbf H}_{0,1}.\]
Note that $\left(A,B\right)$ is not necessarily controllable. For every $i+r=l\ge1$, the matrices ${\mathbf{F}}_{i,r}$ and ${\mathbf{H}}_{i,r}$ are compacted into
\[\begin{array}{l}
{{\mathbf{F}}_l} = \left[ {\begin{array}{*{20}{c}}
{{{\mathbf{F}}_{l,0}}}&{{{\mathbf{F}}_{l - 1,1}}}& \cdots &{{{\mathbf{F}}_{0,l}}}
\end{array}} \right],\\
{{\mathbf{H}}_l} = \left[ {\begin{array}{*{20}{c}}
{{{\mathbf{H}}_{l,0}}}&{{{\mathbf{H}}_{l - 1,1}}}& \cdots &{{{\mathbf{H}}_{0,l}}}
\end{array}} \right].
\end{array}\]

Then we describe the excitation signal. Denote $\sigma\in\mathbb N^+$, $\delta_l={{\mathcal C}^l_{\sigma+l-1}}$. Define ${\mathbf u}\left( v \right):\mathbb R^{\delta_1}\to\mathbb R^m$ as a polynomial function of $v$ of order $L$, i.e.,
\[{\mathbf{u}}\left( v \right) = \sum\limits_{l = 1}^L {{{\mathbf{U}}_l}{v^{\left[ l \right]}}} ,\]
where ${\mathbf U}_l\in \mathbb R^{m\times \delta_l}$ is an unknown matrix. The excitation signal is designed as
\begin{equation}\label{excitation}
\begin{array}{l}
\dot v\left(t\right) = Sv\left(t\right),\\
u\left(t\right) = {\mathbf{u}}\left( v\left(t\right) \right),\\
v\left(0\right)=v_0,
\end{array}
\end{equation}
where
\[S = {\rm{blkdiag}}\left( {0,\left[ {\begin{array}{*{20}{c}}
0&{{\omega _1}}\\
{ - {\omega _1}}&0
\end{array}} \right], \cdots ,\left[ {\begin{array}{*{20}{c}}
0&{{\omega _q}}\\
{ - {\omega _q}}&0
\end{array}} \right]} \right),\]
with $\omega_1,\cdots,\omega_q$ being known positive numbers, $v_0\in\mathbb R^{\delta_1}$ is the initial condition. When $\mathbf U_1\neq0$ and $\mathbf U_l=0, l=2,\cdots,L$, the excitation signal (\ref{excitation}) is a multi-sine wave function with $\omega_i, i=1,\cdots,q$ being the angular frequencies.

Suppose the signals are sampled and denoted by $\left(u_k,y_k,v_k\right)$ for $k=1,\cdots,N$ where $N$ is the total number of samples, $u_k,y_k,v_k$ are nothing but $u\left(t_k\right),y\left(t_k\right),v\left(t_k\right)$ generated by multiple distinct initial conditions $v_0$, for details see \cite{Huang2024Identification}. Then, the input/output signals are corrupted by addictive noise:
\begin{equation}
\begin{array}{*{20}{c}}
\begin{array}{l}
{{\tilde u}_k} = {u_k} + {\mu _k},\\
{{\tilde y}_k} = {y_k} + {\nu _k},
\end{array}&{k = 1, \cdots ,N.}
\end{array}
\end{equation}
where ${\tilde u}_k$ and ${\tilde y}_k$ are the measured signals, ${\mu _k}$ and ${\nu _k}$ are the measurement noise.

\begin{Pro}\label{problem}
Given noise-corrupted samples of the measured input and output data $\{\tilde u_k,\tilde y_k\}$, $k=1,2,\cdots,N$, an integer $\bar n$ as the upper bound for $n$, find, as $N\to\infty$,

\begin{enumerate}
\item{the system order $n$;}

\item{The quadruple $\left(A,B,C,D\right)$ up to a similarity transformation;}

\item{The pair $\left({\mathbf F}_l,{\mathbf H}_l\right)$ for $l=2,3,\cdots,L$, subject to the obtained $\left(A,B,C,D\right)$.}
\end{enumerate}
\end{Pro}

The following assumptions are made for the numerical example.

\begin{Asu}\label{asu_Hurwitz}
$A$ is Hurwitz.
\end{Asu}

\begin{Asu}\label{asu_observable}
$\left(C, A\right)$ is observable.
\end{Asu}

\begin{Asu}\label{asu_controllable_prime}
$\left(A, \left[B, {\mathbf F}_2,\cdots,{\mathbf F}_L\right]\right)$ is controllable.
\end{Asu}

\begin{Asu}\label{asu_bounded}
$n$ has a known upper bound $\bar n$.
\end{Asu}

\section{FDS identification Algorithm}
Consider the following Sylvester equation:
\begin{equation}\label{Sylvester_exam}
\begin{array}{l}
{\mathcal X}{\mathcal S} = {\mathcal A}{\mathcal X} + {\mathcal B}{\mathcal U},\\
{\mathcal Y} = {\mathcal C}{\mathcal X} + {\mathcal D}{\mathcal U},
\end{array}
\end{equation}
where $\mathcal S\in\mathbb R^{\sigma_0\times \sigma_0}$, $\mathcal A\in\mathbb R^{n_0\times n_0}$, $\mathcal B\in\mathbb R^{n_0\times m_0}$,  $\mathcal C\in\mathbb R^{p_0\times n_0}$, $\mathcal D\in\mathbb R^{p_0\times m_0}$, $\mathcal X\in\mathbb R^{n_0\times \sigma_0}$, $\mathcal U\in\mathbb R^{m_0\times \sigma_0}$, $\mathcal Y\in\mathbb R^{p_0\times \sigma_0}$. $\bar n_0$ is an upper bound of $n_0$. We assume that $\mathcal S$ is \emph{diagonalizable}, whose eigenvalues are purely imaginary (including exactly $\alpha$ zero eigenvalues). Moreover, $\mathcal S$ and $\mathcal A$ share no common eigenvalues.

The purpose of the FDS method to find the quintuple $\left(\mathcal A,\mathcal B,\mathcal C,\mathcal D,\mathcal X\right)$ given the input $\left(\mathcal U,\mathcal Y,\mathcal S\right)$. The fairly standard FDS method is summarized in Algorithms \ref{Alg1}-\ref{Alg2}, which are based on Algorithms 2-3 of \cite{Huang2024Identification} and cited here for convenience of the readers. The matrices in Algorithms \ref{Alg1}-\ref{Alg2}  are defined as follows: the input-output data matrices and the extended observability matrix in Algorithms \ref{Alg1}-\ref{Alg2}  are described respectively as
\[{\bar{\mathcal Y}} = \left[ {\begin{array}{*{20}{c}}
\mathcal Y\\
{\mathcal Y\mathcal S}\\
 \vdots \\
{\mathcal Y{\mathcal S^{\bar n_0 - 1}}}
\end{array}} \right],{\bar{\mathcal U}} = \left[ {\begin{array}{*{20}{c}}
\mathcal U\\
{\mathcal U\mathcal S}\\
 \vdots \\
{\mathcal U{\mathcal S^{\bar n_0 - 1}}}
\end{array}} \right],{\mathcal O} = \left[ {\begin{array}{*{20}{c}}
\mathcal C\\
{\mathcal C\mathcal A}\\
 \vdots \\
{\mathcal C{\mathcal A^{\bar n_0 - 1}}}
\end{array}} \right].\]
Moreover, let $\tilde{\mathcal Z}$ stands either for $\tilde{\mathcal X},\tilde{\mathcal U}$ or $\tilde{\mathcal Y}$, then there exists a nonsingular matrix $\mathcal T\in\mathbb R^{\sigma_0\times\sigma_0}$ such that
\[\begin{array}{l}
{{\mathcal T}^{ - 1}}{\mathcal S}{\mathcal T} = {\rm{diag}}\left( {0, \cdots ,0,j{\omega _1}, - j{\omega _1}, \cdots ,j{\omega _q}, - j{\omega _q}} \right),\\
{\mathcal Z}{\mathcal T} = \left[ {{{\tilde {\mathcal Z}}_{0,1}}, \cdots ,{{\tilde {\mathcal Z}}_{0,\alpha }},{{\tilde {\mathcal Z}}_1},\tilde {\mathcal Z}_1^ \star , \cdots ,{{\tilde {\mathcal Z}}_q},\tilde {\mathcal Z}_q^ \star } \right],
\end{array}\]
where $\sigma_0=\alpha+2q$. Finally, define $\mathscr G\left( s \right) = \mathcal C{\left( {sI - \mathcal A} \right)^{ - 1}}\mathcal B + \mathcal D$.

\begin{algorithm}
	\caption{FDS Identification for $\left(\mathcal C,\mathcal A\right)$}
	\label{Alg1}
\textbf{Input:} $\left(\mathcal U,\mathcal Y,\mathcal S\right)$.\quad\textbf{Output:} $\left(\mathcal C,\mathcal A\right)$.

\textbf{Execute}
\begin{enumerate}
\item{Perform the QR decomposition
\begin{eqnarray}\label{QR}
\left[ {\begin{array}{*{20}{c}}
\bar{\mathcal U}^T&\bar{\mathcal Y}^T
\end{array}} \right] = \left[ {\begin{array}{*{20}{c}}
{{Q_1}}&{{Q_2}}
\end{array}} \right]\left[ {\begin{array}{*{20}{c}}
{{R_{1,1}}}&{{R_{1,2}}}\\
O&{{R_{2,2}}}
\end{array}} \right]
\end{eqnarray}
where ${R_{1,1}} \in {\mathbb R^{m_0{\bar n}_0 \times m_0{\bar n}_0}},{R_{2,2}} \in {\mathbb R^{p_0{\bar n}_0 \times p_0{\bar n}_0}}$, ${R_{1,2}} \in {\mathbb R^{m_0{\bar n}_0 \times p_0{\bar n}_0}}$, ${Q_1} \in {\mathbb R^{\sigma_0 \times m_0{\bar n}_0}},{Q_{2}} \in {\mathbb R^{\sigma_0 \times p_0{\bar n}_0}}$;
}

\item{Perform the singular value decomposition (SVD):
\begin{eqnarray}\label{SVD}
R_{2,2}^T = W_{\rm l}\Sigma W_{\rm r}^T,
\end{eqnarray}
where $W_{\rm l},W_{\rm r}\in\mathbb R^{p_0{\bar n}_0\times p_0{\bar n}_0}$ satisfy $W_{\rm l}^TW_{\rm l} =W_{\rm r}^TW_{\rm r} =  I_{p_0{\bar n}_0}$, and $\Sigma={\rm diag}\left(\mu_1,\cdots,\mu_{p_0{\bar n}_0}\right)$ with $\mu_j\ge0,\forall j$ being the singular values;
}

\item{The system order $n_0$ is the number of nonzero eigenvalues of $\Sigma$, and the observability matrix $\mathcal O$ is estimated by the left singular vectors in $W_{\rm l}$ associated with the nonzero singular values;}

\item{The output matrix $\mathcal C$ is the first $p$ rows of ${\mathcal O}$. The system matrix is given by
\begin{eqnarray}\label{shift_property}
\begin{array}{l}
{\cal A} = {\left( {\left[ {{I_{{p_0}\left( {{{\bar n}_0} - 1} \right)}},{O_{{p_0}\left( {{{\bar n}_0} - 1} \right) \times {p_0}}}} \right]{\cal O}} \right)^\dag }\\
 \cdot \left[ {{O_{{p_0}\left( {{{\bar n}_0} - 1} \right) \times {p_0}}},{I_{{p_0}\left( {{{\bar n}_0} - 1} \right)}}} \right]{\cal O}
\end{array}
\end{eqnarray}
by invoking the shift property \cite{Mckelvey1996Subspace}.
}
\end{enumerate}

 \end{algorithm}

\begin{algorithm}
	\caption{FDS Identification for $\left(\mathcal B, \mathcal D, \mathcal X\right)$}
	\label{Alg2}
\textbf{Input:} $\left(\mathcal U,\mathcal Y,\mathcal S, \mathcal A,\mathcal C\right)$.\quad\textbf{Output:} $\left(\mathcal B, \mathcal D, \mathcal X\right)$.

\textbf{Execute}
\begin{enumerate}
\item{Obtain $\left(\mathcal B,\mathcal D\right)$ by solving the linear least-square problem:
\begin{equation}\label{LS}
\begin{array}{l}
\left[ {\begin{array}{*{20}{c}}
{\cal B}\\
{\cal D}
\end{array}} \right] = \mathop {\arg \min }\limits_{\scriptstyle{\cal B} \in {\mathbb R^{{n_0} \times {m_0}}},\hfill\atop
\scriptstyle{\cal D} \in {\mathbb R^{{p_0} \times {m_0}}}\hfill} \sum\limits_{\ell  = 1}^q {{{\left\| {{{\tilde {\cal Y}}_\ell } - {\mathscr G}\left( {j{\omega _\ell }} \right){{\tilde {\cal U}}_\ell }} \right\|}^2}} \\
 + \sum\limits_{\hbar  = 1}^\alpha  {{{\left\| {{{\tilde {\cal Y}}_{0,\hbar }} - {\mathscr G}\left( 0 \right){{\tilde {\cal U}}_{0,\hbar }}} \right\|}^2}} 
\end{array}
\end{equation}

}

\item{Obtain the estimate of $\mathcal X$ with
\begin{equation}\label{solve-X}
\begin{array}{l}
{{\tilde {\cal X}}_\ell } = {\left( {j{\omega _\ell }I - {\cal A}} \right)^{ - 1}}{\cal B}{{\tilde {\cal U}}_\ell },\quad \ell  = 1, \cdots ,q,\\
{{\tilde {\cal X}}_{0,\hbar }} =  - {{\cal A}^{ - 1}}{\cal B}{{\tilde {\cal U}}_{0,\hbar }},\quad \hbar  = 1, \cdots ,\alpha ,
\end{array}
\end{equation}
and the identity
\[{\mathcal X} = \left[ {{{\tilde {\mathcal X}}_{0,1}}, \cdots ,{{\tilde {\mathcal X}}_{0,\alpha }},{{\tilde {\mathcal X}}_1},\tilde {\mathcal X}_1^ \star , \cdots ,{{\tilde {\mathcal X}}_q},\tilde {\mathcal X}_q^ \star } \right]{\mathcal T}^{-1}.\]

}

\end{enumerate}

\end{algorithm}

\section{Simulations}

The numerical example of a second-order PNS is studied:
\begin{equation}\label{simulation_examp}
\begin{array}{l}
{{\dot x}_1} =  - {x_1} - u - 5x_2^2,\\
{{\dot x}_2} =  - 2{x_2} + 0.3x_1^2 + 3{x_1}{x_2},\\
{y_1} =  - 6{x_1},\\
{y_2} = 3{x_2},
\end{array}
\end{equation}
which can be written in the form of Eq. (\ref{eq_preliminary_nldequ_x0}) with 
\begin{equation}\label{Uctrb_examp}
\begin{array}{l}
A = \left[ {\begin{array}{*{20}{c}}
{ - 1}& 0\\
0&{ - 2}
\end{array}} \right],B = \left[ {\begin{array}{*{20}{c}}
{ - 1}\\
0
\end{array}} \right],C = \left[ {\begin{array}{*{20}{c}}
{ - 6}& 0\\
0&{ 3}
\end{array}} \right],\\
{{\mathbf F}_{2,0}} = \left[ {\begin{array}{*{20}{c}}
0&0&{ - 5}\\
{0.3}&3&0
\end{array}} \right].
\end{array}
\end{equation}
It is assumed to be known that $D$, ${\mathbf F}_{1,1}$, ${\mathbf F}_{0,2}$, ${\mathbf H}_2$ are all zero matrices of appropriate dimensions. Moreover, it is assumed that $\bar n=3$, $L=2$, which are known, too. It can be verified that $A$ is Hurwitz, $\left(C,A\right)$ is observable and $\left(A,[B,{\mathbf F}_{2,0}]\right)$ is controllable, but $\left(A,B\right)$ is uncontrollable. The excitation signal contains five frequency components which are $\omega_1=0.13,\omega_2=0.79,\omega_3=2.65,\omega_4=7.81,\omega_5=18.37$ rad/s. ${\mathbf U}_1=0.05\left([1,2,4,8,16]\otimes [1,0]\right)$ while ${\mathbf U}_k=0$ for every $k\ge 2$. 
 
Now we start describing the identification process. Note that the following procedures are not exactly copying Algorithm 4 of \cite{Huang2024Analytical} which is designed for general PNSs. But we believe when identifying a nonlinear system, it is beneficial to exploit \emph{a priori} information about the underlying system as much as possible, not only to simplify the SID algorithm, but also to expect better performance.

In the simulation, we compare the SID performance of the CM-based method either when noise is absent or present. When noise is absent and $\left({\mathbf U}_l,{\mathbf Y}_l\right), l=1,2,3,4$ is accurately known, by running Algorithms \ref{Alg1}-\ref{Alg2} with $\left({\mathbf U}_1,{\mathbf Y}_1,S\right)$ as input for the following Sylvester equation (see \cite{Huang2024Analytical} for its derivation)
\begin{equation}\label{Sylvester_lPR}
\begin{array}{l}
{{\mathbf{X}}_{1,{\rm{c}}}}S = {A_{1,\rm c}}{{\mathbf{X}}_{1,{\rm{c}}}}{\rm{ + }}{B_{1,\rm c}}{{\mathbf{U}}_1},\\
{{\mathbf{Y}}_1} = {C_{1,\rm c}}{{\mathbf{X}}_{1,{\rm{c}}}} + D{{\mathbf{U}}_1}.
\end{array}
\end{equation}
where ${\mathbf{X}}_{1,{\rm{c}}}\in\mathbb R^{n_{1,\rm c}\times\delta_1}$, 
$A_{1,\rm c}\in\mathbb R^{n_{1,\rm c}\times n_{1,\rm c}}$, $C_{1,\rm c}\in\mathbb R^{p\times n_{1,\rm c}}$, $B_{1,\rm c}\in\mathbb R^{n_{1,\rm c}\times m}$,
one obtains the singular values $1.2,1.9\times 10^{-14},1.4\times 10^{-15},2.4\times 10^{-32},0,0$. Then it is clear that $n_{1,\rm c}=1$, and ${\mathbf X}_{1,\rm c}$ is obtained as part of the output of Algorithm \ref{Alg2}. Now consider the Sylvester equation
\begin{equation}\label{Sylv-4th-V}
\begin{array}{l}
{{{\bar{\mathbf X}}}_2}{{\bar S}_2} = A{{{\bar{\mathbf X}}}_2} + \left[ {\begin{array}{*{20}{c}}
B&{{{\mathbf{F}}_{2,0}}}
\end{array}} \right]{{\mathcal R}^\prime_2}{{{\bar{\mathbf V}}}''_{2}},\\
{{{\bar{\mathbf Y}}}_2} = C{{{\bar{\mathbf X}}}_2}.
\end{array}
\end{equation}
where ${{{\bar{\mathbf X}}}_2}=[{\mathbf X}_1,{\mathbf X}_2]$, ${{\bar S}_2}={\rm blkdiag}\left(S,{S^{\left\langle 2 \right\rangle }}\right)$, ${{{\mathcal R}}^\prime_2} = {\rm{blkdiag}}\left( {1,{\mathcal R}_2^{2,0}} \right)$ (since ${\mathcal R}_2^{2,0}$ is not used in the SID algorithm, its expression is omitted, for details the readers are referred to \cite{Huang2024Analytical}),
\[{{{\bar{\mathbf V}}}''_{2}} = \left[ {\begin{array}{*{20}{c}}
{{{\mathbf{U}}_1}}&{{{\mathbf{U}}_2}}\\
O&{{\mathbf{V}}_2^{2,0}}
\end{array}} \right],\] 
in which
\[{\mathbf{V}}_2^{2,0} = {\left[ {\begin{array}{*{20}{c}}
{{{\mathbf{X}}_{1,{\rm{c}}}}}\\
O_{\left(\bar n-n_{1,{\rm c}}\right)\times \delta_1}
\end{array}} \right]^{\left( 2 \right)}}N_2^\sigma. \]
Since ${\mathbf X}_{1,\rm c}$ is obtained, ${{{\bar{\mathbf V}}}''_{2}}$ can be calculated. Run Algorithm \ref{Alg1} with $\left({{{\bar{\mathbf V}}}''_{2}}, {{\bar{\mathbf Y}}}_2, {\bar S}_2\right)$ as input (${{{\bar{\mathbf V}}}''_{2}}$ needs to be transformed into a matrix of full row rank, see \cite{Huang2024Analytical} for details) for Eq. (\ref{Sylv-4th-V}) to obtain the estimate of $\left(C,A\right)$ as 
\[
\hat A = \left[ {\begin{array}{*{20}{c}}
{ - 1.0000}&{0}\\
{ 0}&{ - 2.0000}
\end{array}} \right],\hat C = \left[ {\begin{array}{*{20}{c}}
{0.5774}&{ 0}\\
{ 0}&{ - 0.2182}
\end{array}} \right].\]
Next, ${\mathbf X}_1,{\mathbf X}_2,{\mathbf X}_3$ are obtained one by one by running Algorithm \ref{Alg2} with $\left({\mathbf U}_1,{\mathbf Y}_1,S,\hat A,\hat C\right)$, $\left({\mathbf Z}_2,{\mathbf Y}_2,{S^{\left\langle 2 \right\rangle }},\hat A,\hat C\right)$ and $\left({\mathbf Z}_3^\prime,{\mathbf Y}_3,{S^{\left\langle 3 \right\rangle }},\hat A,\hat C\right)$ as input (${\mathbf Z}_2$ and ${\mathbf Z}_3^\prime$ need to be transformed into matrices of full row rank, see \cite{Huang2024Analytical} for details), respectively, where
\[\begin{array}{l}
{{{{\mathbf Z}_2}}} = {\left[ {{\mathbf{U}}_2^{\rm{T}},{{\left( {{{\mathbf Z}}_2^{2,0}} \right)}^{\rm{T}}},{{\left( {{{\mathbf Z}}_2^{1,1}} \right)}^{\rm{T}}},{{\left( {{{\mathbf Z}}_2^{0,2}} \right)}^{\rm{T}}}} \right]^{\rm{T}}},\\
{{{{\mathbf Z}'_3}}} = \left[ {{\mathbf{U}}_3^{\rm{T}},{{\left( {{{\mathbf Z}}_2^{2,0}} \right)}^{\rm{T}}},{{\left( {{{\mathbf Z}}_2^{1,1}} \right)}^{\rm{T}}},{{\left( {{{\mathbf Z}}_2^{0,2}} \right)}^{\rm{T}}}} \right.\\
{\left. {,{{\left( {{{\mathbf Z}}_3^{2,0}} \right)}^{\rm{T}}},{{\left( {{{\mathbf Z}}_3^{1,1}} \right)}^{\rm{T}}},{{\left( {{{\mathbf Z}}_3^{0,2}} \right)}^{\rm{T}}}} \right]^{\rm{T}}},
\end{array}\]
in which
\[\begin{array}{l}
{\mathbf{Z}}_2^{2,0} = M_2^n{\mathbf{X}}_1^{\left( 2 \right)}N_2^\sigma ,\\
{\mathbf{Z}}_2^{1,1} = \left( {{{\mathbf{X}}_1} \otimes {{\mathbf{U}}_1}} \right)N_2^\sigma ,\\
{\mathbf{Z}}_2^{0,2} = M_2^m{\mathbf{U}}_1^{\left( 2 \right)}N_2^\sigma ,\\
{\mathbf{Z}}_3^{2,0} = M_2^n\left[ {{{\mathbf{X}}_1} \otimes \left( {{{\mathbf{X}}_2}M_2^\sigma } \right) + \left( {{{\mathbf{X}}_2}M_2^\sigma } \right) \otimes {{\mathbf{X}}_1}} \right]N_3^\sigma ,\\
{\mathbf{Z}}_3^{1,1} = \left[ {{{\mathbf{X}}_1} \otimes \left( {{{\mathbf{U}}_2}M_2^\sigma } \right) + \left( {{{\mathbf{X}}_2}M_2^\sigma } \right) \otimes {{\mathbf{U}}_1}} \right]N_3^\sigma ,\\
{\mathbf{Z}}_3^{0,2} = M_2^m\left[ {{{\mathbf{U}}_1} \otimes \left( {{{\mathbf{U}}_2}M_2^\sigma } \right) + \left( {{{\mathbf{U}}_2}M_2^\sigma } \right) \otimes {{\mathbf{U}}_1}} \right]N_3^\sigma .
\end{array}\]
Finally, consider
\begin{equation}\label{Sylv-4th-Z}
\begin{array}{l}
{{{\bar{\mathbf X}}}_4}{{\bar S}_4} = A{{{\bar{\mathbf X}}}_4} + \left[ {\begin{array}{*{20}{c}}
B&{{{\mathbf{F}}_{2,0}}}
\end{array}} \right]{{{\bar{\mathbf Z}}}''_4},\\
{{{\bar{\mathbf Y}}}_4} = C{{{\bar{\mathbf X}}}_4},
\end{array}
\end{equation}
where
\[{{{\bar{\mathbf Z}}}''_4} = \left[ {\begin{array}{*{20}{c}}
{{{\mathbf{U}}_1}}&{{{\mathbf{U}}_2}}&{{{\mathbf{U}}_3}}&{{{\mathbf{U}}_4}}\\
O&{{{\mathbf Z}}_2^{2,0}}&{{{\mathbf Z}}_3^{2,0}}&{{{\mathbf Z}}_4^{2,0}}
\end{array}} \right],\]
in which
\[\begin{array}{l}
{\mathbf{Z}}_4^{2,0} = M_2^n\left[ {{{\mathbf{X}}_1} \otimes \left( {{{\mathbf{X}}_3}M_3^\sigma } \right) + {{\left( {{{\mathbf{X}}_2}M_2^\sigma } \right)}^{\left( 2 \right)}}} \right.\\
\left. { + \left( {{{\mathbf{X}}_3}M_3^\sigma } \right) \otimes {{\mathbf{X}}_1}} \right]N_4^\sigma .
\end{array}\]
It is checked that the associated PE condition for $\left(\left[ {\begin{array}{*{20}{c}}
B&{{{\mathbf{F}}_{2,0}}}
\end{array}} \right],0,\bar{\mathbf X}_4\right)$ of Eq. (\ref{Sylv-4th-Z}) is satisfied (see \cite{Huang2024Analytical} for details). Then it is valid to run Algorithm \ref{Alg2} with $\left({{\bar{\mathbf Z}}}''_4,{{\bar{\mathbf Y}}}_4,{\bar S}_4,\hat A,\hat C\right)$ as input for Eq. (\ref{Sylv-4th-Z}) given that $D$, ${\mathbf F}_{1,1}$, ${\mathbf F}_{0,2}$, ${\mathbf H}_2$ are all zero, to obtain the estimate of $\left(B,{\mathbf F}_{2,0}\right)$ as part of the output as follows:
\[\hat B = \left[ {\begin{array}{*{20}{c}}
{10.392}\\
{ 0}
\end{array}} \right],{{\hat{\mathbf F}}_{2,0}} = \left[ {\begin{array}{*{20}{c}}
{0}&{ 0}&{-0.2749}\\
{-0.0382}&{ 0.2887}&{0}
\end{array}} \right].\]

Then we consider the case when noise is present. The noise signals at the input and output channels are mutually independent zero-mean Gaussian white noise with variance $0.03$, respectively, resulting a SNR around $80$dB at both channels. The sampling intervals at both channels are the same and equal to $0.01$s. In the presence of noise, a total number of $50$ experiments are run, each of which admits a distinct realization of the noise. In each experiment, $10^5$ time-domain samples are used to run Algorithm 1 of \cite{Huang2024Identification} to estimate $\left({\mathbf U}_l,{\mathbf Y}_l\right)$ for $l=1,2,3,4$. The average SID error over the $50$ experiment is calculated.

After $\left({\mathbf U}_l,{\mathbf Y}_l\right), l=1,2,3,4$ is estimated, two SID algorithms are used in the noise-corrupted case, the first (Method I) is exactly the one described in the noiseless case. The only difference between Methods I and II is the algorithm for estimating ${\mathbf X}_1,{\mathbf X}_2$ and ${\mathbf X}_3$. With Method II, consider the equation
\begin{equation}\label{simu_syl4}
\begin{array}{l}
{{{\bar{\mathbf X}}}_{4}}{{\bar S}_{4}} = A{{{\bar{\mathbf X}}}_{4}} + \left[ {\begin{array}{*{20}{c}}
B&{{{\mathbf{F}}_{2,0}}}
\end{array}} \right]{\mathcal R}_4''{{{\bar{\mathbf V}''}_4}},\\
{{{\bar{\mathbf Y}}}_{4}} = C{{{\bar{\mathbf X}}}_{4}} ,
\end{array}
\end{equation}
where ${\mathcal R}_4''  = {\rm{blkdiag}}\left( {1,{\mathcal R}_2^{2,0},{\mathcal R}_3^{2,0},{\mathcal R}_4^{2,0}} \right)$ (since ${\mathcal R}_2^{2,0}$ ${\mathcal R}_3^{2,0}$, ${\mathcal R}_4^{2,0}$ are not used in the SID algorithm, their expression is omitted, for details the readers are referred to \cite{Huang2024Analytical}),
\[{{{\bar{\mathbf V}''}}_4} = \left[ {\begin{array}{*{20}{c}}
{{{\mathbf{U}}_1}}&{{{\mathbf{U}}_2}}&{{{\mathbf{U}}_3}}&{{{\mathbf{U}}_4}}\\
O&{{\mathbf{V}}_2^{2,0}}&O&O\\
O&O&{{\mathbf{V}}_3^{2,0}}&O\\
O&O&O&{{\mathbf{V}}_4^{2,0}}
\end{array}} \right],\]
in which
\[\begin{array}{l}
{\mathbf{V}}_3^{2,0} = \left[ \begin{array}{l}
\left[ {\begin{array}{*{20}{c}}
{{{\mathbf{X}}_{1,{\rm{c}}}}}\\
O
\end{array}} \right] \otimes \left( {\left[ {\begin{array}{*{20}{c}}
{{{\mathbf{X}}_{2,{\rm{c}}}}}\\
O
\end{array}} \right]M_2^\sigma } \right)\\
\left( {\left[ {\begin{array}{*{20}{c}}
{{{\mathbf{X}}_{2,{\rm{c}}}}}\\
O
\end{array}} \right]M_2^\sigma } \right) \otimes \left[ {\begin{array}{*{20}{c}}
{{{\mathbf{X}}_{1,{\rm{c}}}}}\\
O
\end{array}} \right]
\end{array} \right]N_3^\sigma ,\\
{\mathbf{V}}_4^{2,0} = \left[ {\begin{array}{*{20}{c}}
{\left[ {\begin{array}{*{20}{c}}
{{{\mathbf{X}}_{1,{\rm{c}}}}}\\
O
\end{array}} \right] \otimes \left( {\left[ {\begin{array}{*{20}{c}}
{{{\mathbf{X}}_{3,{\rm{c}}}}}\\
O
\end{array}} \right]M_3^\sigma } \right)}\\
{{{\left( {\left[ {\begin{array}{*{20}{c}}
{{{\mathbf{X}}_{2,{\rm{c}}}}}\\
O
\end{array}} \right]M_2^\sigma } \right)}^{\left( 2 \right)}}}\\
{\left( {\left[ {\begin{array}{*{20}{c}}
{{{\mathbf{X}}_{3,{\rm{c}}}}}\\
O
\end{array}} \right]M_3^\sigma } \right) \otimes \left[ {\begin{array}{*{20}{c}}
{{{\mathbf{X}}_{1,{\rm{c}}}}}\\
O
\end{array}} \right]}
\end{array}} \right]N_4^\sigma ,
\end{array}\]
Note that ${\mathbf X}_{2,\rm c},{\mathbf X}_{3,\rm c}$ can be obtained similarly to ${\mathbf X}_{1,\rm c}$ and therefore the algorithms are omitted. ${{{\bar{\mathbf X}}}_{4}}=[{\mathbf X}_1,{\mathbf X}_2,{\mathbf X}_3,{\mathbf X}_4]$ is obtained alternatively by running Algorithm \ref{Alg2} for Eq. (\ref{simu_syl4}) with $\left({{{\bar{\mathbf V}}}''_{4}},{{{\bar{\mathbf Y}}}_{4}},\bar S_4,\hat A,\hat C\right)$ as input  (${{{\bar{\mathbf V}}}''_{4}}$ needs to be transformed into a matrix of full row rank, see \cite{Huang2024Analytical} for details).

\begin{figure}
\centering
\includegraphics[width=0.52\textwidth,height=0.3\textwidth]{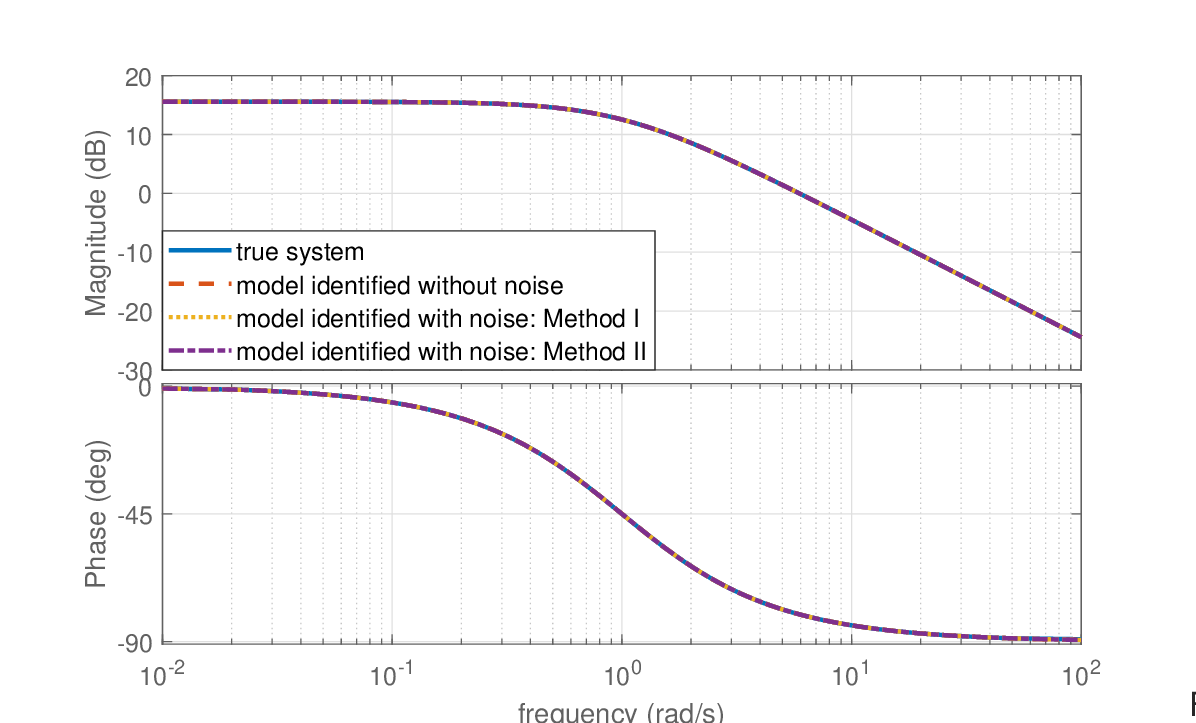}
\caption{Bode diagram for $\mathcal G_{1,1,1}\left(s\right)$ with or without measurement noise. The solid blue line is for the true system, the dashed red line is for the model identified without noise, the dotted yellow line is for the model identified with noise by Method I, the dotted-dashed purple line is for the model identified with noise by Method II. The color of the lines has the same meaning in the subsequent figures. In this figure, all lines are overlapped due to negligible ID errors.}
\label{fig-G111}
\end{figure}

\begin{figure}
\centering
\includegraphics[width=0.52\textwidth,height=0.3\textwidth]{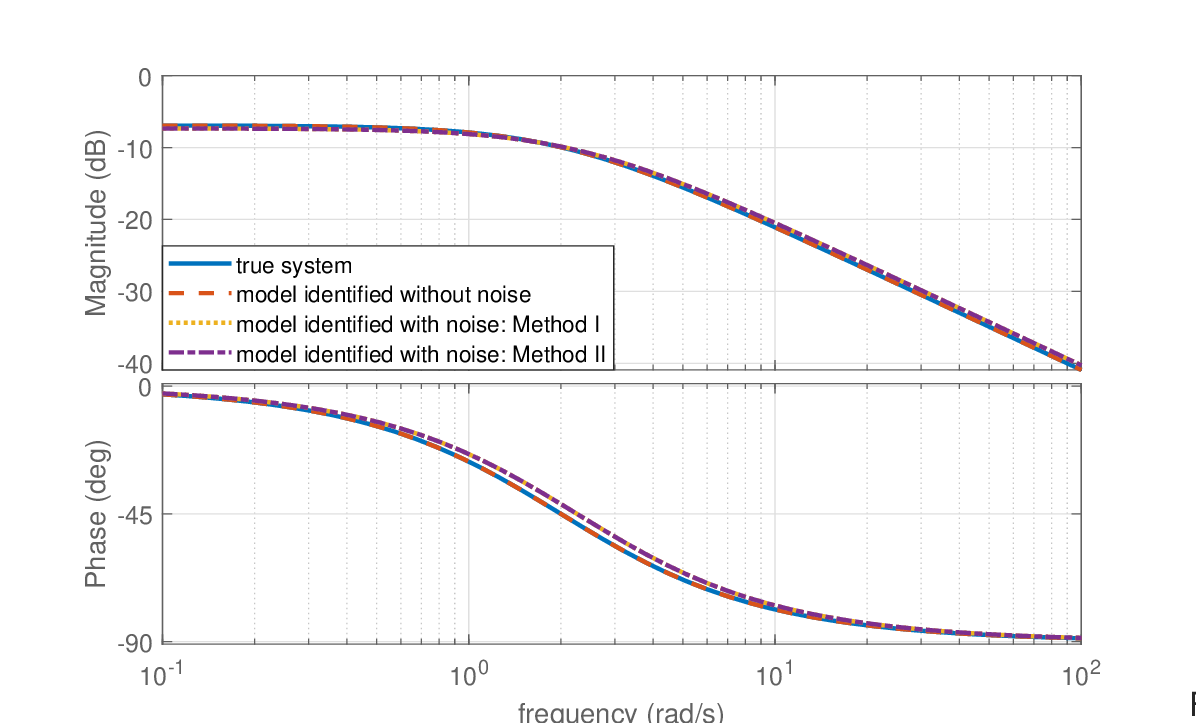}
\caption{Bode diagram for $\mathcal G_{2,2,1}\left(s\right)$ identified with or without measurement noise. The model identified without noise is overlapped with the true system, while the model identified with noise has a small ID error.}
\label{fig-G221}
\end{figure}

\begin{figure}
\centering
\includegraphics[width=0.52\textwidth,height=0.3\textwidth]{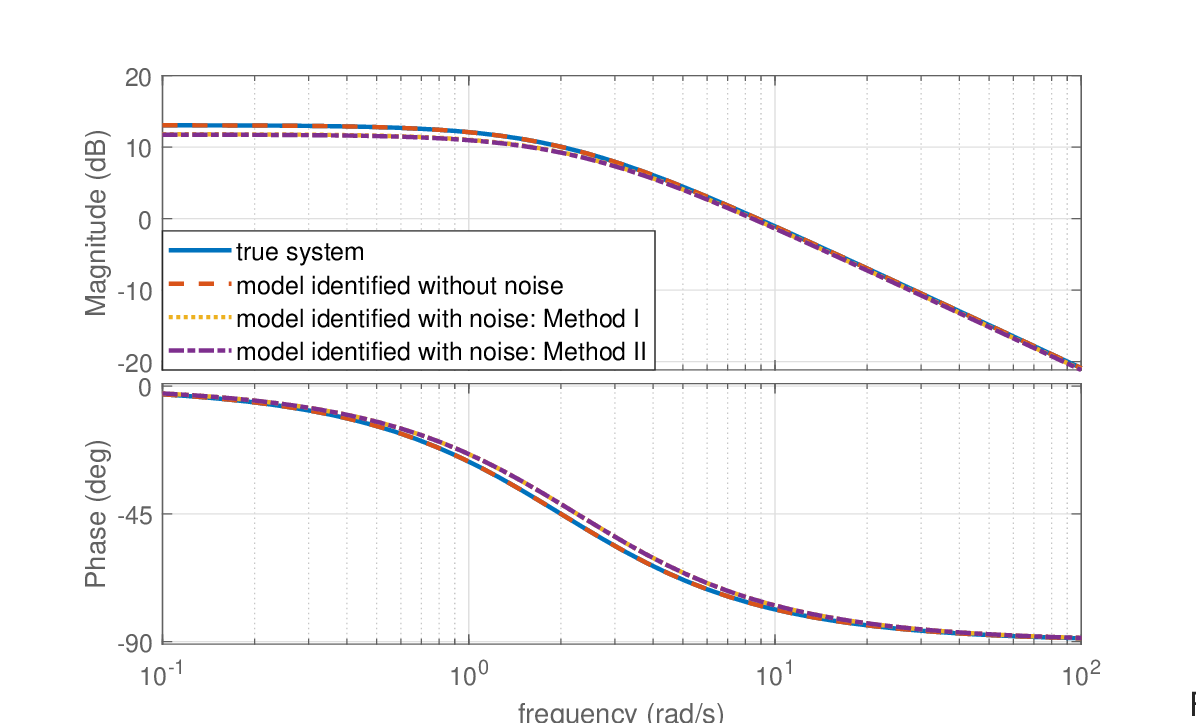}
\caption{Bode diagram for $\mathcal G_{2,2,2}\left(s\right)$ identified with or without measurement noise. The model identified without noise is overlapped with the true system, while the model identified with noise has a small ID error.}
\label{fig-G222}
\end{figure}

\begin{figure}
\centering
\includegraphics[width=0.52\textwidth,height=0.3\textwidth]{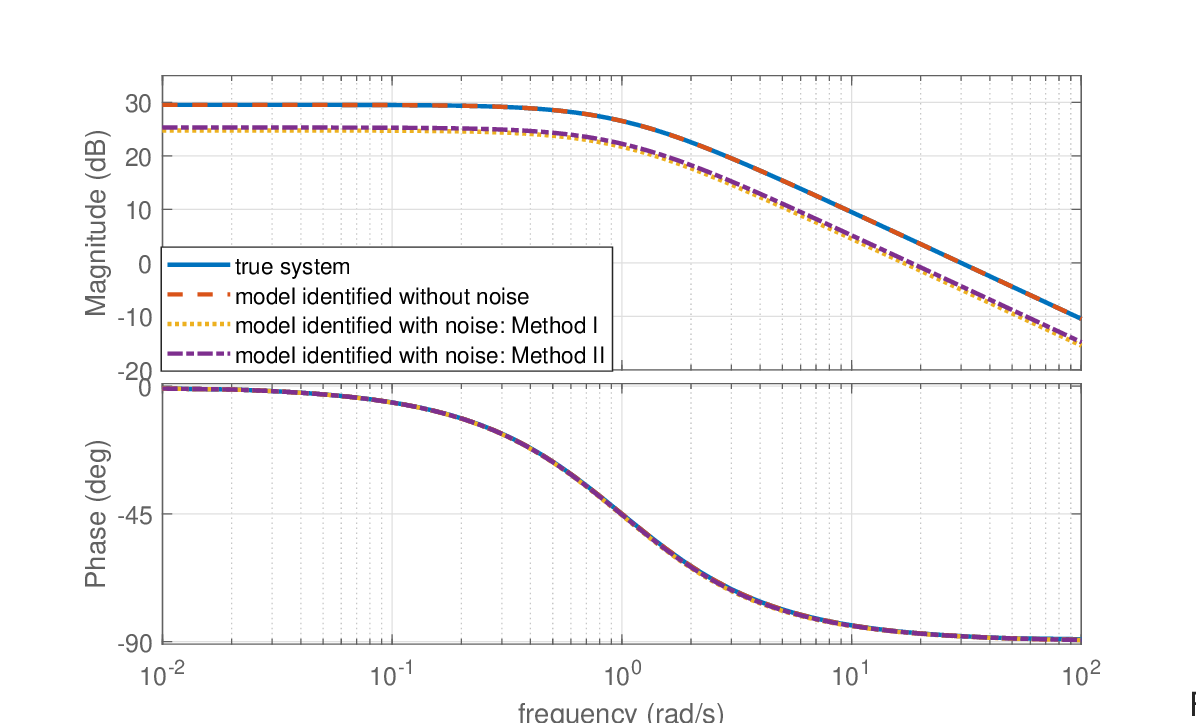}
\caption{Bode diagram for $\mathcal G_{2,1,3}\left(s\right)$ identified with or without measurement noise. The model identified without noise is overlapped with the true system, while the model identified with noise has a small ID error.}
\label{fig-G213}
\end{figure}

\begin{figure}
\centering
\includegraphics[width=0.52\textwidth,height=0.3\textwidth]{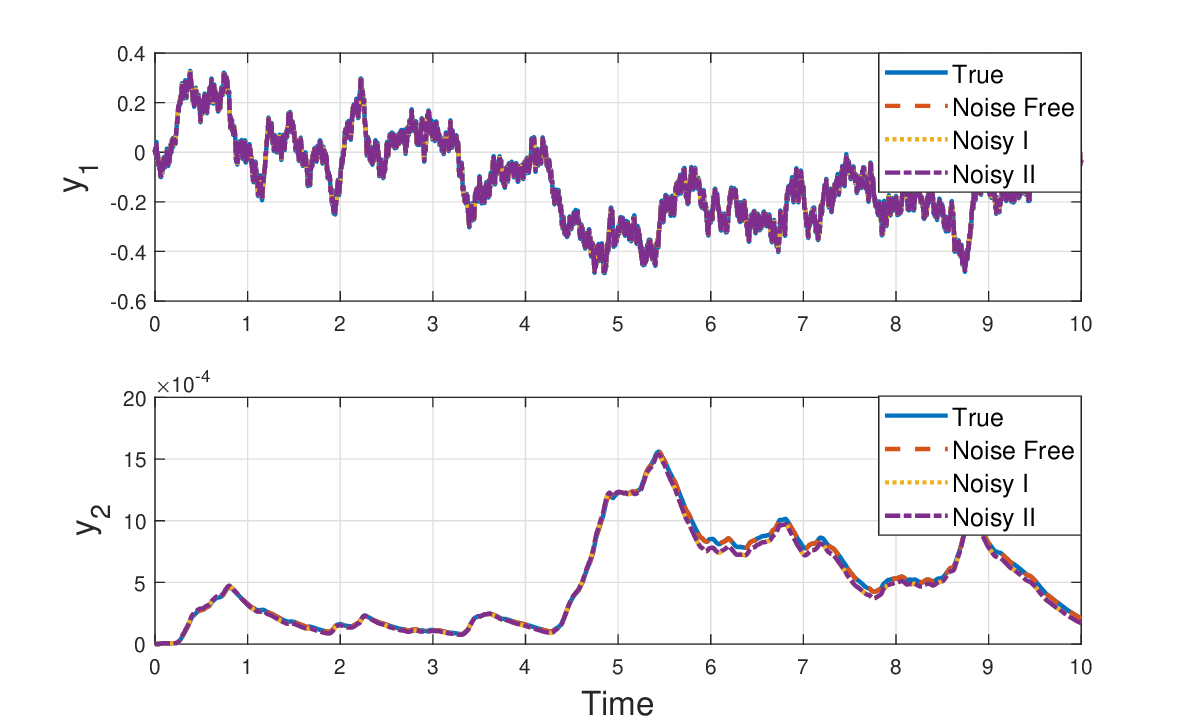}
\caption{The output of the system (\ref{simulation_examp}) and the identified models driven by the same discrete-time white noise.} 
\label{fig-y}
\end{figure}

Let
\[\begin{array}{l}
{\mathcal G_1}\left( s \right) = C{\left( {sI - A} \right)^{ - 1}}B\in\mathbb C^{2\times 1},\\
{\mathcal G_2}\left( s \right) = C{\left( {sI - A} \right)^{ - 1}}{{\mathbf{F}}_{2,0}} \in\mathbb C^{2\times 3},
\end{array}\]
in which all entries are zero except the ones in Row $1$, Column $1$ of ${\mathcal G_1}\left( s \right)$, Row $2$, Column $1$, Row $2$, Column $2$ and Row $1$, Column $3$ of ${\mathcal G_2}\left( s \right)$. They are denoted by $\mathcal G_{1,1,1}\left( s \right)$, $\mathcal G_{2,2,1}\left( s \right)$, $\mathcal G_{2,2,2}\left( s \right)$, and $\mathcal G_{2,1,3}\left( s \right)$, respectively. The identification error is measured by the following norm ratio:
\begin{equation}\label{Ratio}
\frac{{{{\left\| {{\mathcal G}_\Delta\left( s \right) - \hat {\mathcal G}_\Delta\left( s \right)} \right\|}_\infty }}}{{{{\left\| {{\mathcal G}_\Delta\left( s \right)} \right\|}_\infty }}},
\end{equation}
where $\left\|\cdot\right\|_\infty$ is the $H_\infty$ norm \cite{Zhou1995Robust}, ${{{\mathcal G}_\Delta}\left( s \right)}$ stands for certain entry of the true transfer function, while ${\hat {\mathcal G}_\Delta\left( s \right)}$ is the estimate of ${{{\mathcal G}_\Delta}\left( s \right)}$. It is known that $\mathcal G_1\left(s\right)$ is invariant under the transformation from the coordinate framework (CF) of the true system to that of the identified model, but $\mathcal G_2\left(s\right)$ is not. So the system and the model should be compared in the common CF (see \cite{Huang2024Identification} for the CF conversion formula for $m=1$). In the simulation, we use the CF of the true system. The SID errors and bode diagrams for the true as well as identified $\mathcal G_1\left( s \right),\mathcal G_2\left( s \right)$ by Methods I and II are shown in Table I and Figs. \ref{fig-G111}-\ref{fig-G213}. It is shown that, in this example, SID error is negligible if noise is absent, and is within a reasonable range when noise is present. Moreover, Method I and II have almost the same average performance as far as this example is concerned. The output of the system (\ref{simulation_examp}) and the identified models driven by the same discrete-time white noise is shown in Fig. \ref{fig-y}.

\renewcommand{\arraystretch}{1.5} 
\begin{table}[tp]
  \centering
  \fontsize{6.5}{8}\selectfont
 
  \caption{Identification Errors Measured By Eq.(\ref{Ratio}).}
  \label{tab}
    \begin{tabular}{cccc}
    \hline
    Transfer Function &  ID Error (noise free)& ID Error (noisy, I)& ID Error (noisy, II)\cr
    \hline
    $\mathcal G_{1,1,1}\left(s\right)$ & $4.26\times10^{-16}$&$8.44\times10^{-6}$&$8.44\times10^{-6}$\cr
    \hline
    $\mathcal G_{2,2,1}\left(s\right)$&$2.14\times10^{-15}$&$4.66\times10^{-2}$&$4.66\times10^{-2}$\cr
    \hline
    $\mathcal G_{2,2,2}\left(s\right)$&$4.14\times10^{-15}$&$1.34\times10^{-1}$&$1.40\times10^{-1}$\cr
    \hline
    $\mathcal G_{2,1,3}\left(s\right)$&$4.57\times10^{-13}$&$4.24\times10^{-1}$&$3.84\times10^{-1}$\cr
    \hline

    \end{tabular}

\end{table}

\ifCLASSOPTIONcaptionsoff
  \newpage
\fi

\end{document}